\documentclass[runningheads]{llncs}
\usepackage[T1]{fontenc}
\usepackage{graphicx}
\usepackage{booktabs}
\usepackage[misc]{ifsym}
\usepackage{amssymb,amsmath}
\usepackage{caption}
\usepackage{longtable}
\usepackage{multirow}
\usepackage{array}
\usepackage{algorithm,algorithmic}
\usepackage{textcomp}
\usepackage{stfloats}
\usepackage{url}
\usepackage{verbatim}
\usepackage{svg}
\usepackage{adjustbox}
\usepackage{bm}
\usepackage{lineno}                  
\usepackage[colorlinks=true,linkcolor=black,citecolor=blue,urlcolor=blue]{hyperref}

\newcommand{\corr}{(\Letter)}

\usepackage{mwe}

\begin{document}

\title{Diffusion-augmented Graph Contrastive Learning for Collaborative Filter}

\titlerunning{Diffusion-augmented Graph Contrastive Learning for Collaborative Filter}

\author{Fan Huang \and
Wei Wang\corr }

\authorrunning{F. Huang and W. Wang.}

\institute{School of Data Science and Engineering, East China Normal University, China\\
\email{51265903093@stu.ecnu.edu.cn},
\email{wwang@dase.ecnu.edu.cn}}

\maketitle              

\begin{abstract}
Graph-based collaborative filtering has been established as a prominent approach in recommendation systems, leveraging the inherent graph topology of user-item interactions to model high-order connectivity patterns and enhance recommendation performance. Recent advances in Graph Contrastive Learning (GCL) have demonstrated promising potential to alleviate data sparsity issues by improving representation learning through contrastive view generation and mutual information maximization. However, existing approaches lack effective data augmentation strategies.  Structural augmentation risks distorting fundamental graph topology, while feature-level perturbation techniques predominantly employ uniform noise scales that fail to account for node-specific characteristics. To solve these challenges, we propose Diffusion-augmented Contrastive Learning (DGCL), an innovative framework that integrates diffusion models with contrastive learning for enhanced collaborative filtering. Our approach employs a diffusion process that learns node-specific Gaussian distributions of representations, thereby generating semantically consistent yet diversified contrastive views through reverse diffusion sampling.  DGCL facilitates adaptive data augmentation based on reconstructed representations, considering both semantic coherence and node-specific features.  In addition, it explores unrepresented regions of the latent sparse feature space, thereby enriching the diversity of contrastive views. Extensive experimental results demonstrate the effectiveness of DGCL on three public datasets.

\keywords{Diffusion Model \and Graph Contrast Learning \and Collaborative Filtering.}
\end{abstract}
\section{Introduction}
Collaborative Filtering (CF) remains a cornerstone of recommendation systems, aiming to predict user preferences by leveraging historical interactions between users and items~\cite{a25,a26}. Traditional approaches predominantly focus on node embedding representation learning such as matrix factorization (MF)~\cite{a3} or graph-based methods like DeepWalk~\cite{a1}. Recently the advent of graph neural networks (GNNs), particularly graph convolutional networks (GCN)~\cite{a2}, has revolutionized this landscape through recursive neighborhood aggregation. GNN-based CF models like NGCF~\cite{a4} and LightGCN~\cite{a5} explicitly encode multi-hop connectivity via message-passing mechanisms, thereby modeling indirect collaborative effects beyond immediate interactions.

Despite these advances, graph-based collaborative filtering grapples with data sparsity, which induces suboptimal node representations that fail to generalize beyond observed interactions. To mitigate this, graph contrastive learning (GCL)~\cite{a23,a24,a30} has emerged as a promising paradigm, introducing self-supervised signals to alleviate popularity bias and enhance generalization. For instance, SimGCL~\cite{a6} proposes perturbing node embeddings with uniform random noise to regulate the uniformity of representation distributions and mitigate popularity bias. Existing GCL frameworks generally adopt two augmentation paradigms: structural augmentation (e.g., node/edge dropout, subgraph sampling) and feature augmentation (e.g., feature masking, noise injection, or clustering).

However, conventional augmentation strategies suffer from two fundamental limitations. Structural perturbations risk disrupting vital topological dependencies, such as distorting key nodes or user-item interactions, while feature-level augmentations like uniform noise addition, disregard the heterogeneous semantic contexts of nodes and unique features~\cite{a27}, applying identical distortion scales to both high-degree items and long-tail entities. This approach generates semantically inconsistent contrastive views, ultimately degrading representation learning. Consequently, designing augmentation strategies that preserve critical graph semantics while diversifying contrastive views remains an open challenge.

To address this, we propose a Diffusion-augmented Graph Contrastive Learning (DGCL) framework, which integrates diffusion-based probabilistic modeling to generate adaptive augmentation views. Specifically, DGCL estimates node-specific Gaussian distributions over embeddings, enabling adaptive sampling of augmented views conditioned on each node's latent semantics.  The forward diffusion progressively injects controlled noise, while the reverse denoising process generates semantically consistent yet diverse contrastive views. This mechanism synthesizes high-quality augmented contrastive views that maintain semantic coherence and node-specific features in a node-adaptive manner.
Furthermore, the diffusion process inherently unlocks the hidden value in sparse interaction data, allowing the model to explore unrepresented regions of the feature space without distorting graph topology, enriching the diversity and quality of contrastive pairs. 
Therefore, DGCL not only preserves structural integrity but also tailors augmentation granularity to individual nodes, addressing the augmentation limitation of conventional GCL methods. The experiment analysis demonstrates that DGCL effectively balances augmentation robustness and semantic coherence, advancing the state-of-the-art in graph-based collaborative filtering. The main contributions of this paper are summarized as follows:
\begin{itemize}
\item[$\bullet$] We propose a novel Diffusion-augmented Graph Contrastive Learning (DGCL) framework which incorporates the generative diffusion model into contrastive learning for collaborative filtering.
\item[$\bullet$] We design a data augmentation scheme, in which the diffusion model was used to generate high-quality contrastive views that consider semantic correlation and node-specific features. 
\item[$\bullet$]Extensive experiments on three public datasets confirm the superior effectiveness of the DGCL.

\end{itemize}

\section{Related Work}
\subsection{Graph Contrastive Learning based Recommendation}
Contrastive learning has emerged as a pivotal paradigm in Self-supervised learning (SSL)~\cite{a24} to address data sparsity and popularity bias in recommendation systems. LightGCN~\cite{a5} pioneered a streamlined approach by eliminating non-linear transformations in graph convolution, emphasizing high-order connectivity through pure neighborhood aggregation.  To inject robustness, SGL~\cite{a7}  introduced structural augmentation via stochastic node or edge dropout, contrasting multiple subgraph views through a shared GNN encoder to learn invariant user or item representations. SimGCL~\cite{a6} advanced this paradigm by perturbing embeddings with uniform noise. This approach further alleviated popularity bias and smoothly adjusted the uniformity of learned representations. NCL~\cite{a8} explicitly modeled high-order semantics by contrasting nodes with neighborhood prototypes, improving representation alignment and uniformity. Meanwhile, DirectAU~\cite{a9} directly optimized embedding alignment and uniformity without augmentation, achieving strong performance in sparse scenarios. Although these methods highlight the versatility of GCL,   the challenge of data augmentation and constructing sophisticated contrastive views remain unresolved.

\subsection{Diffusion based Recommendation}
The diffusion model is the most popular approach in image generation with the development of the denoising diffusion probabilistic model (DDPM)~\cite{a10} and score-based generative model~\cite{a11}. Recently, advancements in diffusion models have inspired novel paradigms for addressing key challenges in recommendation systems and early approaches focused on interaction modeling. DiffRec~\cite{a12} pioneered diffusion for recommendation by reformulating user-item interactions as a denoising sequence, where personalized signals are preserved through noise injection and reconstruction. Its two extensions demonstrated diffusion’s versatility in balancing scalability and preference specificity. 
CF-Diff~\cite{a13} advanced DiffRec by explicitly encoding high-order user-item graph connectivity through multi-step denoising, iteratively refining node representations to uncover latent collaborative patterns beyond first-order interactions. Recent innovations have diversified diffusion’s role in the recommendation.
BSPM~\cite{a14} introduced an image-inspired paradigm, blurring (noise injection) and sharpening (denoising) user-item interactions to distill noise-robust collaborative signals.
DiffKG~\cite{a16} synergized diffusion models with knowledge graph learning, leveraging the semantic information from knowledge graphs to enhance recommendation quality. 
By defining a generalized diffusion process on the item-item similarity graph, GiffCF~\cite{a15} effectively modeled the distribution of user-item interactions, achieving superior performance in collaborative filtering tasks. 
Other methods combine the diffusion model and contrastive learning, but they are not in the realm of collaborative filtering. DiffGCL~\cite{a28} was employed in graph classification, and DiffCL~\cite{a29} was proposed for the multimodal recommendation. Different from the above diffusion-based recommendation system, our DGCL model focuses on data augmentation and generates semantically consistent yet diverse contrastive views in collaborative filtering.

\section{Preliminaries}
Collaborative filtering aims to recommend some relevant items to users based on implicit interaction. Given the user set $\mathcal{U}=\{u\}$, items set $\mathcal{I}=\{i\}$ and the implicit interaction matrix $ \mathcal{R}\in\{0,1\}^{|\mathcal{U}|\times |\mathcal{I}|}$, the recommendation model predicts the potential interactions between user and item. Specifically, the graph-based collaborative filtering treats the user set and item set as nodes and abstracts the interaction matrix as the bipartite graph $\mathcal{G}=\{\mathcal{V},\mathcal{E}\}$, where $\mathcal{V}=\mathcal{U} \cup \mathcal{I}$ represents the node set and $\mathcal{E}=\{(u, i)|u\in \mathcal{U},i\in \mathcal{I}, \mathcal{R}_{u, i}=1\}$ represents the edge set.
LightGCN is the most common backbone to aggregate the high-order neighbors' information in the recommendation system. Given the initial embedding $e_u$ and $e_i$, representing the embeddings of user $u$ and item $i$, the basic aggregation process is:

\begin{align}
    e_u^{(l+1)}=\sum_{i\in \mathcal{N}_u}\frac{1}{\sqrt{|\mathcal{N}_u||\mathcal{N}_i|}}e_i^{(l)},e_i^{(l+1)}=\sum_{u\in \mathcal{N}_i}\frac{1}{\sqrt{|\mathcal{N}_i||\mathcal{N}_u|}}e_u^{(l)},
\end{align}

where $\mathcal{N}_u$ and $\mathcal{N}_i$ represent the neighbor sets of user $u$ and item $i$, respectively. The final embeddings are obtained by averaging all layer outputs:
\begin{align}
    \mathbf{e}_{u}=\frac{1}{L} \sum_{l=1}^{L} \mathbf{e}_{u}^{(l)}, \mathbf{e}_{i}=\frac{1}{L} \sum_{l=1}^{L} \mathbf{e}_{i}^{(l)}.
\end{align}




\section{Methods}
\begin{figure}
\noindent\includegraphics[width=\textwidth]{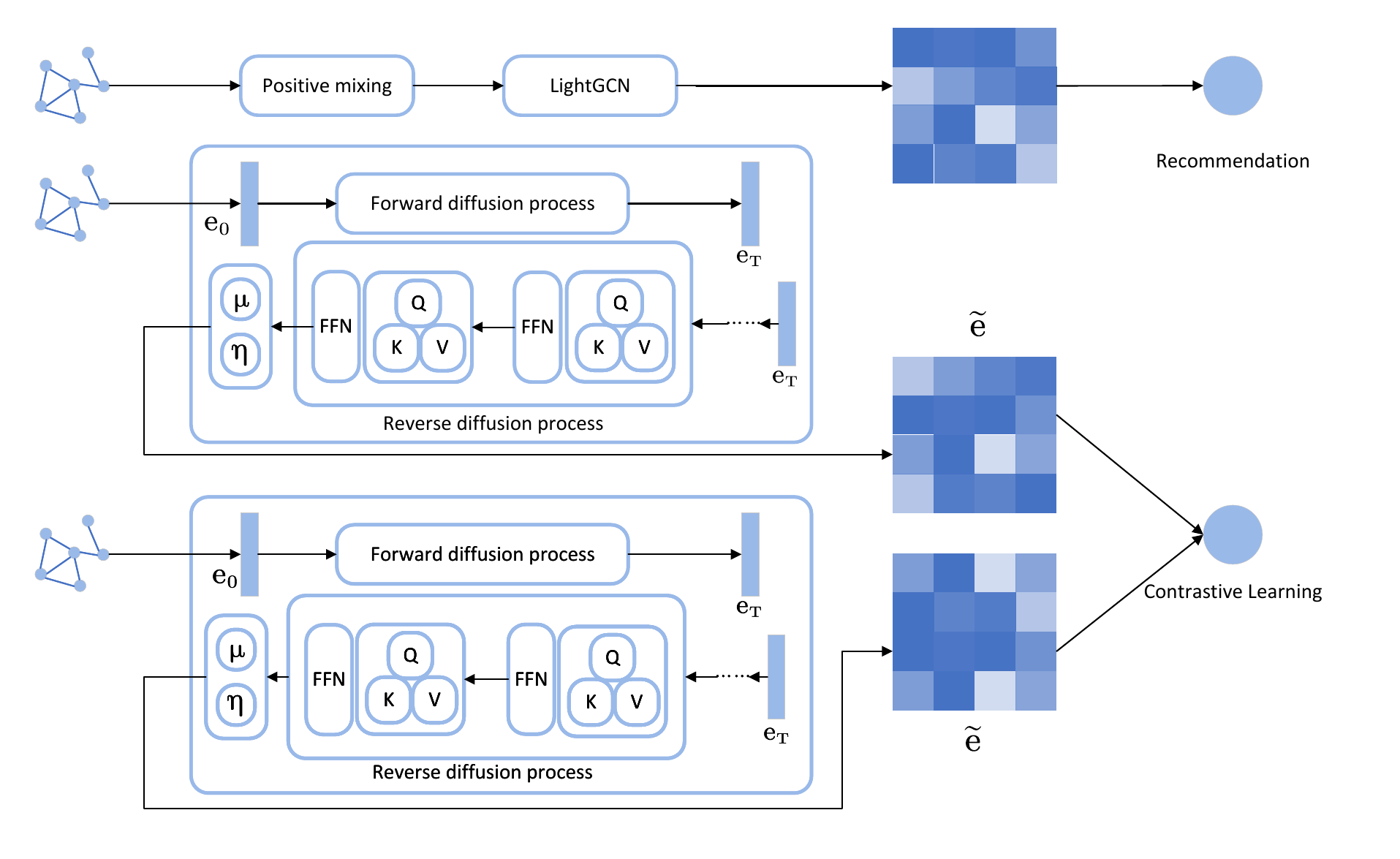}
\caption{Overall framework of DGCL. }
\label{diffusion_contrastive}
\end{figure}

As illustrated in Figure \ref{diffusion_contrastive}, DGCL consists of three modules: graph collaborative relation learning, diffusion augmentation module and contrastive learning module. In graph collaborative relation learning, we adopt the basic GNN encoder to capture the high-order neighbors and latent embedding representation. Specifically, we implement LightGCN to aggregate feature aggregation and integrate a novel negative sampling strategy to enhance discriminative capability. In the diffusion augmentation module, we incorporate dual diffusion processes that systematically inject Gaussian noise through forward diffusion steps, subsequently generating contrastive views by sampling from the learned posterior distributions. Finally, the contrastive learning module establishes cross-view comparisons between users and items through a dual-channel architecture. The framework is optimized via a joint training objective combining collaborative filtering and contrastive losses.
\subsection{Graph Collaborative Relation Learning}
Given the initial embedding, $e_u$, $e_i$ denotes the representation of the user $u$ and item $i$. The core idea of LightGCN is to learn node embeddings by message propagation and information aggregation, iteratively stacking convolution layers. Inspired by~\cite{a18}, we adopt positive mixing to enhance the negative sample quality. The main logic is to generate a more indistinguishable synthetic negative sample by injecting the embedding information of the positive sample into the candidate negative sample. The positive mixing operation is formalized as follows:
\begin{align}
    \mathbf{e}_{i^{-}}^{\prime(l)}=\alpha^{(l)} \mathbf{e}_{i^{+}}^{(l)}+\left(1-\alpha^{(l)}\right) \mathbf{e}_{i^{-}}^{(l)}, \alpha^{(l)} \in(0,1),
\end{align}
where $ \mathbf{e}_{i^{-}}^{(l)}$ denotes the negative item representation in the $l$ layer, $ \mathbf{e}_{i^{+}}^{(l)}$ denotes the positive item representation in the $l$ layer and $\mathbf{e}_{i^{-}}^{\prime(l)}$ denotes the synthetic negative item representation in the $l$ layer. By injecting positive sample information, the synthesized negative sample is closer to the positive sample in the embedding space, forcing the model to learn finer differentiation ability, thus improving the accuracy of recommendation ranking and providing high-quality hard negative samples. Additionally, each layer carries out mixing operations independently so that the synthesized negative samples are interfered at different levels, which comprehensively challenges the discriminant logic of the model. For more details, please refer to \cite{a18}.

\subsection{Diffusion Augmentation Module}
While conventional graph diffusion methods typically inject noise into adjacency matrices to produce interaction matrices, our approach operates in the latent embedding space to generate augmented contrastive views. In general, the module consists of two coordinated phases: a forward noise injection process and a parameterized reverse generation process.

\noindent\textbf{Forward Diffusion Process.}
Given nodes representation $e$, including the user representation $e_u$ and item representation $e_i$, we progressively inject the Gaussian noise through $T$ diffusion steps governed by a variance schedule $\{\beta_t\}_{t=1}^{T}$. 
 The forward process forms a Markov chain defined by:
\begin{align}
    q\left(e_{1: T} \mid x_{0}\right)=\prod_{t=1}^{T} q\left(e_{t} \mid e_{t-1}\right),
\end{align}
where $e_0$ represents the original node embedding and $e_T$ denotes the noised node representation. $q\left(e_{t} \mid e_{t-1}\right)$ denote the conditional probability distribution of noise injection in the forward diffusion process and the specific condition distribution for each step is:
\begin{align}
    q\left(e_{t} \mid e_{t-1}\right)=\mathcal{N}\left(e_{t} ; \sqrt{1-\beta_{t}} e_{t-1}, \beta_{t} \mathbf{I}\right),
\label{forward}
\end{align}
where $\beta$ controls the Gaussian noise scales at each time step $t$, $ \mathcal{N}$ refers to the Gaussian distribution. By exploiting the reparameterization trick~\cite{a17},
the noised representation can be expressed as $\boldsymbol{e}_{t}=\sqrt{\bar{\alpha}_{t}} \boldsymbol{e}_{0}+\sqrt{1-\bar{\alpha}_{t}} \epsilon_{t}$, where $\alpha_{t}=1-\beta_{t},\bar{\alpha}_{t}=\prod_{t^{\prime}=1}^{t}\alpha_{t^{\prime}},\mathrm{~and~}\epsilon\sim\mathcal{N}(0,I)$. $t$ refers to the diffusion step. $e_t$ approaches a standard Gaussian distribution if $t$ is sufficiently large. Consequently, noise-perturbed user and item embeddings are generated to train the model, thereby facilitating the estimation of Gaussian parameters.

\noindent\textbf{Reverse Diffusion Process.}
The reverse process generates data by denoising step by step, it is also defined as a parameterized Markov chain, and each step satisfies the Gaussian process:
\begin{align}
    p_\theta(e_{t-1}|e_t)=\mathcal{N}
\begin{pmatrix}
e_{t-1};\mu_\theta(e_t,t),\sigma_t^2\mathbf{I}
\end{pmatrix},
\label{reverse}
\end{align}
where $\mu_\theta, \sigma_t$ are the learned parameters in the model. Specifically, we employ the 
two-layers Transformer~\cite{a31} architecture as the encoder, including the multi-head attention and feed-forward network. First, to encode the time step in the diffusion process, the time coding employs the improved sinusoidal position coding:
\begin{align}
    PE(t,2i)=\sin(\frac{t}{10000^{2i/d}}),\quad PE(t,2i+1)=\cos(\frac{t}{10000^{2i/d}}).
\end{align}
The time encoding is incorporated by the feature space through a feature-wise linear modulation~\cite{a19}, which can be formulated as:
\begin{align}
    \gamma,\eta=\text{TimeMLP}(t), h=(\gamma+1)e+\eta
\end{align}
Subsequently, the denoised user and item representations are derived via the attention mechanism and feed-forward network. Specifically,
\begin{align}
    h^{(l+1)}=\mathrm{LayerNorm}(h^{(l)}+\mathrm{Multi\_Attention}(h^{(l)}))\\
    \mathrm{Attention}(Q,K,V)=\mathrm{softmax}\left(\frac{QK^T}{\sqrt{d_k}}\right)V,
\end{align}
where $Q=W_Qh,K=W_Kh,V=W_Vh$. Hence, we can predict the learned Gaussian noise parameters to fit the node representation in the diffusion process. 

\noindent\textbf{Diffusion Loss.}
The diffusion augmentation module learns the parameters of the Gaussian distribution governing the latent user and item embeddings $e_0$. Unlike conventional approaches that directly predict noise, DGCL is designed to recover the underlying embedding structure from the original feature space. The learning objective is twofold: (1) to minimize the Kullback-Leibler (KL) divergence between the learned and target distributions, and (2) to maximize the evidence lower bound (ELBO) of the observed embeddings:
\begin{align}
    \mathcal{L}_{diff}(\theta) = \mathbb{E}_{t, e_0} \left[ \left\| e_0 - f_{\theta}(e_t, t) \right\|^2 \right],
\label{diffusion_loss}
\end{align}
where $f_{\theta}$ is the predicted augmented representation mapping function and the Training pseudo-code is represented in Algorithm~\ref{diffuison_train}

\noindent\textbf{Contrastive View Inference.}
The embedding encoder $f(x_t,t)$, derived from the aforementioned diffusion process, enables the generation of augmented contrastive view embeddings. Our model leverages the estimated distribution parameters to iteratively approximate the posterior distribution $p_\theta(e_{t-1}|e_t)$, thereby inferring refined representations. The posterior mean and covariance of $p_\theta(e_{t-1}|e_t)$ are calculated as follows:
\begin{align}
    \mu_{\theta}(e_t, t) &= \frac{1}{\sqrt{\alpha_t}} \left( e_t - \frac{\beta_t}{\sqrt{1-\bar{\alpha}_t}} \epsilon_{\theta}(e_t, t) \right),\\
    \sigma_{t}^{2}(t) &= \frac{1 - \bar{\alpha}_{t-1}}{1 - \bar{\alpha}_{t}} \beta_{t}.
    \label{infer}
\end{align}
Therefore, we can generate iteratively augmented contrastive view embedding $\widetilde{e}$, including the user contrastive view $\widetilde{e}_u$ and item contrastive view $\widetilde{e}_i$ :
\begin{align}
    e_{t-1} = \frac{1}{\sqrt{\alpha_t}} \left( e_t - \frac{\beta_t}{\sqrt{1 - \bar{\alpha}_t}} \epsilon_{\theta}(e_t, t) \right) + \sigma_t z,
    \label{et-1}
\end{align}
where $z \sim \mathcal{N}(0, \mathbf{I})$. This process generates high-quality synthetic contrastive views through augmentation, which share semantic similarity in the feature space while preserving personalized characteristics for individual node embeddings. Additionally, it uncovers unrepresented embeddings in the feature space, mitigating data sparsity and enhancing the diversity of contrastive views. The inference procedure is outlined in Algorithm~\ref{diffuison_infer}

\begin{algorithm}[!h]
    \caption{Algorithm of DGCL Training}
    \renewcommand{\algorithmicrequire}{\textbf{Input:}}
    \renewcommand{\algorithmicensure}{\textbf{Output:}}
    
    \begin{algorithmic}[1]
        \REQUIRE \small user and item embedding $E$ and randomly initialized $\theta$. 
        \ENSURE optimized $ \theta $.  
        
        \STATE  Sample a batch of node embedding $e\in E$.
        \WHILE{converged}
            \STATE Sample $ t \sim \mathcal{U}(1,T)$ , $ \epsilon \sim \mathcal{N}(1,I)$;
            \STATE Compute the noised embedding $e_t$ given the $e_0$,$\epsilon$ via Eq.~\ref{forward};
            \STATE Predict the noise from $e_t$ iteratively by Eq.~\ref{reverse};
            \STATE Calculate the loss $\mathcal{L}_{diff}$ according to the Eq.~\ref{diffusion_loss};
            \STATE Take the gradient descent step on $ \nabla_{\theta} \mathcal{L}_{t}$ to optimize $\theta$;
        \ENDWHILE
    \end{algorithmic}
    \label{diffuison_train}
\end{algorithm}

\begin{algorithm}[!h]
    \caption{Algorithm of DGCL Inference}
    \renewcommand{\algorithmicrequire}{\textbf{Input:}}
    \renewcommand{\algorithmicensure}{\textbf{Output:}}
    
    \begin{algorithmic}[2]
        \REQUIRE \small embedding prediction model $f_{\theta}$ from the diffusion process, node embedding. $e_0$
        \ENSURE the augmented contrastive view embedding $\widetilde{e}$, including the user contrastive view $\widetilde{e}_u$ and item contrastive view $\widetilde{e}_i$.  
        
        \STATE  Sample Gaussian noise $z \in \mathcal{N}(0,I)$.
        \STATE  Compute the initial noised data $e_t$ in Eq.~\ref{forward}.
        \FOR{$t=T,\dots,1$}
            \STATE Calculate the $\mu_{\theta}(e_t, t)$ and $e_{t-1}$ via Eq.~\ref{infer} and Eq.~\ref{et-1}.
        \ENDFOR
    \end{algorithmic}
    \label{diffuison_infer}
\end{algorithm}

\subsection{Contrastive Learning Module}
In DGCL, we train two distinct diffusion augmentation modules to generate diverse augmented contrastive views for users and items. Each augmented embedding is tailored to the original feature space of its corresponding node. The user and item contrastive losses are defined as:
\begin{align}
    \mathcal{L}_{cl}^{U}=\sum_{u\in\mathcal{B}_{u}}-\mathrm{log}\frac{\exp(\widetilde{\mathbf{e}}_{\mathbf{u}}^{\prime}{}^{T}\widetilde{\mathbf{e}}_{\mathbf{u}}^{\prime\prime}/\tau_{1})}{\sum_{j\in\mathcal{B}_{u}}\exp(\widetilde{\mathbf{e}}_{\mathbf{u}}^{\prime}{}^{T}\widetilde{\mathbf{e}}_{\mathbf{j}}^{\prime\prime}/\tau_{1})},\\\mathcal{L}_{cl}^{I}=\sum_{i\in\mathcal{B}_{i}}-\mathrm{log}\frac{\exp(\widetilde{\mathbf{e}}_{\mathbf{i}}^{\prime T}\widetilde{\mathbf{e}}_{\mathbf{i}}^{\prime\prime}/\tau_{1})}{\sum_{j\in\mathcal{B}_{i}}\exp(\widetilde{\mathbf{e}}_{\mathbf{i}}^{\prime T}\widetilde{\mathbf{e}}_{\mathbf{j}}^{\prime\prime}/\tau_{1})},
    \label{cl}
\end{align}

where $\tau_{1}$ is the temperature hyperparameter in the contrastive loss which scales the similarity scores (dot products) between embeddings.
\subsection{Model Training}
The DGCL consists of two independent training processes: one is to train the diffusion process for the augmented contrastive view representation. The optimal objective is presented in Eq.~\ref{diffusion_loss}. Another is to optimize the recommendation and contrastive module, which adopts the joint learning strategy. It is formulated as follows:

\begin{align}
\mathcal{L}_{joint}=\mathcal{L}_{rec}+\lambda\mathcal{L}_{cl},
\end{align}
where the $L_{rec}$ denotes the recommendation loss BPR: 
\begin{align}
\mathcal{L}_{r e c}=-\log \left(\sigma\left(\mathbf{e}_{u}^{\top} \mathbf{e}_{i}-\mathbf{e}_{u}^{\top} \mathbf{e}_{j}\right)\right) ,
\end{align}
and $L_{cl}$ denotes the contrastive loss which includes the $\mathcal{L}_{cl}^{U}$ and $\mathcal{L}_{cl}^{I}$ in the Eq.~\ref{cl}.

\section{Experiments}
In this section, we conduct extensive experiments to evaluate the performance of DGCL on three benchmark datasets and analyze the key module of the model.
\subsection{Experimental Setup}
\textbf{Datasets.} The experience employs three public datasets in different scenarios. (1) Douban-Book~\cite{a24} originates from the Chinese social platform Douban, including user ratings of books, reviews and social relationships between users. (2) Gowalla\footnote{http://snap.stanford.edu/data/loc-gowalla.html}~\cite{a9} is a location-based social network dataset and records the user behavior and friend relationship. (3) Amazon-Kindle~\cite{a24} focuses on customer purchases and rating data from the Amazon Kindle E-book Store, including rich product attributes and user behavior interactions.

\noindent \textbf{Baseline.} We compare DGCL with other competitive baselines, mainly including three categories: (1)GNN based models: MF~\cite{a3}, LightGCN~\cite{a5}; (2) Graph contrastive learning based models: SGL~\cite{a7}, SimGCL~\cite{a6}, DirectAU~\cite{a9}, NCL~\cite{a8}; (3) other self-supervised learning based models: BUIR~\cite{a21}, SSL4Rec~\cite{a22}, SelfCF~\cite{a20}. 

\noindent \textbf{Experimental Settings.}
the LighGCN is employed  as the basic recommendation embedding. The hidden dimension and learning rate of the DGCL are searched from \{64, 128, 256, 512, 1024\} and\{1e-2, 1e-3, 4e-4, 1e-4\}, respectively. The number of GNN layers is selected from \{1,2,3\}. $\lambda$ is searched in \{ 0.01, 0.2, 0.3\} and timestep of diffusion is searched in \{10,20,30,50\}. The noise $\beta$ is tined in range of \{1e-5, 2e-2\}.
In performance metrics, we adopt the widely used ranking metrics to evaluate the model, including the Recall@K (R@K) and the NDCG@K (N@K), where $K \in \{10,20\}$.

\begin{table}[]
\caption{DGCL Performance Comparison with different methods on three datasets.}
\adjustbox{max width=\textwidth}{
\begin{tabular}{c|cccc|cccc|cccc}
\toprule
\multirow{2}{*}{Models} & \multicolumn{4}{c|}{Douban-Book}   & \multicolumn{4}{c|}{Gowalla}      & \multicolumn{4}{c}{Amazon-Kindle}    \\
 & R@10    & N@10   & R@20   & N@20   & R@10   & N@10   & R@20   & N@20   & R@10    & N@10   & R@20   & N@20    \\
 \midrule

BPR-MF                  & 0.0869  & 0.0949 & 0.1296 & 0.1045 & 0.1158 & 0.0833 & 0.1695 & 0.0988 & 0.10873 & 0.0801 & 0.14949 & 0.0923 \\
LighGCN                 & 0.1042  & 0.1195 & 0.1516 & 0.1278 & 0.1362 & 0.0876 & 0.1976 & 0.1152 & 0.1425  & 0.1063 & 0.1906  & 0.1208  \\
SGL                     & 0.1103  & 0.1357 & 0.1551 & 0.1419 & 0.1255 & 0.1371 & 0.1783 & 0.1517 & 0.1445  & 0.1054 & 0.1974  & 0.12138 \\
NCL                     & 0.1121  & 0.1377 & 0.1576 & 0.1439 & 0.1272 & 0.1384 & 0.181 & 0.1535 & 0.1384  & 0.1005 & 0.1867  & 0.1152  \\
BUIR                    & 0.0640  & 0.0736 & 0.1036 & 0.0824 & 0.0842 & 0.0940 & 0.1216 & 0.1040 & 0.0551  & 0.0373 & 0.0830  & 0.0458  \\
SSL4Rec                 & 0.0811  & 0.0849 & 0.1142 & 0.0926 & 0.0576 & 0.0508 & 0.0958 & 0.0649 & 0.1491  & 0.1152 & 0.1924  & 0.1283  \\
SelfCF                  & 0.0595  & 0.0662 & 0.0944 & 0.0741 & 0.0798 & 0.0909 & 0.1146 & 0.0998 & 0.0403  & 0.0269 & 0.0642  & 0.0341  \\
DirectAU                & 0.0999  & 0.1136 & 0.1365 & 0.1197 & 0.1091 & 0.1144 & 0.1584 & 0.1295 & 0.1225  & 0.0882 & 0.1757  & 0.1041  \\
SimiGCL                 & 0.1218  & 0.1470 & 0.1731 & 0.1540 & 0.1279 & 0.1391 & 0.1823 & 0.1544 & 0.1449  & 0.1067 & 0.1967  & 0.1222  \\
DGCL                 & $\textbf{0.1292}$ & $\bm{0.1593} $&$\bm{ 0.1782}$ & $\bm{0.1639}$ & $\bm{0.1307}$ &$ \bm{0.1424}$ & $\bm{0.1855}$ & $\bm{0.1577} $& $\bm{0.1495}$ & $\bm{0.1090}$ & $\bm{0.2052}$  & $\bm{0.1259}$ \\
\bottomrule
\end{tabular}}
\label{result}
\end{table}

\subsection{Experimental Results}
The comparative performance of DGCL against baseline methods is summarized in Table~\ref{result}. DGCL achieves superior performance across three distinct public datasets in collaborative filtering for recommendation systems, demonstrating the efficacy of its diffusion-augmented approach. Notably, the method outperforms traditional noise injection strategies by generating perturbations that better preserve semantic correlations in the feature space. Specially, On the Douban-Book dataset, DGCL attains an N@10 score of 15.93\% and N@20 score of 16.39\%, improving 1.23\% and 0.99\% respectively over the strongest baseline, SimGCL. For Amazon-Kindle, DGCL surpasses SimGCL by 0.85\% in R@20. Uniform noise may introduce uncorrelated perturbations in the embedded space, disproportionately widening differences between sample pairs and degrading semantic coherence. This hinders contrastive learning from capturing meaningful signals. In contrast, DGCL’s diffusion process generates enhanced contrastive samples through iterative denoising, enabling more feature-adaptive embedding representations that maintain semantic consistency while diversifying contrastive views. The superior performance verifies that DGCL can generate high-quality contrastive augmented views then SimGCL.

In addition, SGL randomly discards edges or nodes, risking structural degradation and poor performance across all datasets. NCL marginally outperforms SGL and is approximately 0.2\% higher in Douban-Book. However, it is constrained by prototype quality and may be limited by the clustering quality. If the clustering is inaccurate, the prototype may not effectively represent the user's interests, resulting in a decreased comparison effect. For instance, its inferior performance on Amazon-Kindle likely stems from challenges in clustering extensive and heterogeneous commodity data, whereas DGCL flexibly generates diverse augmentations without relying on clustering.
Therefore, DGCL provides a suitable data augmentation that combines semantic correlation in feature space and unique features for each node representation to avoid semantic deviation caused by random noise perturbations. It not only maintains the graphs' topological characteristics but also uncovers unrepresented representations in the feature space through progressive generation.

\begin{table}[]
\caption{Ablation study of DGCL, DGCL-w/o diff denotes the model variant without diffusion augmentation, and DGCL-w/o neg represents the variant without negative sampling.}
\adjustbox{max width=\textwidth}{
\begin{tabular}{c|cccc|cccc|cccc}
\toprule
\multirow{2}{*}{Models} & \multicolumn{4}{c|}{Douban-Book}  & \multicolumn{4}{c|}{Gowalla}      & \multicolumn{4}{c}{Amazon-Kindle} \\
                        & R@10   & N@10   & R@20   & N@20   & R@10   & N@10   & R@20   & N@20   & R@10   & N@10   & R@20   & N@20   \\ \hline
DGCL - w/o diff         & 0.1246 & 0.1562 & 0.1740 & 0.1574 & 0.1294 & 0.1413 & 0.1845 & 0.1486 & 0.1486 & 0.1082 & 0.2042 & 0.1250 \\
DGCL - w/o neg          & 0.0796 & 0.0928 & 0.1251 & 0.1015 & 0.0896 & 0.0994 & 0.1411 & 0.1223 & 0.0629 & 0.0784 & 0.1643 & 0.1462 \\
DGCL                    & $\bm{0.1292}$ & $\bm{0.1593}$ & $\bm{0.1782}$ & $\bm{0.1639}$ & $\bm{0.1307}$ & $\bm{0.1424}$ & $\bm{0.1855}$ & $\bm{0.1577}$ & $\bm{0.1495}$ & $\bm{0.1090}$ & $\bm{0.2052}$ & $\bm{0.1259}$\\
\bottomrule
\end{tabular}}
\label{ablation}

\end{table}
\begin{table}[]
\caption{DGCL performance on different graph inference layer L.}
\adjustbox{max width=\textwidth}{
\begin{tabular}{c|cccc|cccc|cccc}
\toprule
\multirow{2}{*}{Models} & \multicolumn{4}{c|}{Douban-Book}  & \multicolumn{4}{c|}{Gowalla}      & \multicolumn{4}{c}{Amazon-Kindle} \\
                        & R@10   & N@10   & R@20   & N@20   & R@10   & N@10   & R@20   & N@20   & R@10   & N@10   & R@20   & N@20   \\ \hline
DGCL layer=1            & 0.1269 & 0.1545 & 0.1742 & 0.1588 & 0.1287 & 0.1402 & 0.1818 & 0.1549 & 0.1488 & 0.1002 & 0.2023 & 0.1243 \\
DGCL layer=2            & 0.1279 & 0.1583 & 0.1777 & 0.1633 & 0.1296 & 0.1416 & 0.1849 & 0.1571 & 0.1494 & 0.1090 & 0.2051 & 0.1258 \\
DGCL layer=3            & $\bm{0.1292}$ & $\bm{0.1593}$ & $\bm{0.1782}$ & $\bm{0.1638 }$& $\bm{0.1307} $& $\bm{0.1424}$ &$ \bm{0.1855}$ & $\bm{0.1577}$ &$ \bm{0.1495}$ & $\bm{0.1090}$ & $\bm{0.2052} $& $\bm{0.1259}$\\
\bottomrule
\end{tabular}}
\label{layer}
\end{table}

\subsection{Ablation Study}
In this section, we conduct ablation studies to evaluate the contributions of two key modules: negative sampling and diffusion augmentation. The experimental results on three public datasets are presented in Table~\ref{ablation}. Here, DGCL-w/o diff denotes the model variant without diffusion augmentation, and DGCL-w/o neg represents the variant without negative sampling. 
The metrics exhibit a noticeable decline when diffusion augmentation is removed, with reductions of 0.31\% in N@10 and 0.65\% in N@20 on the Douban-Book dataset. This underscores the module’s ability to uncover latent representations in sparse data distributions, effectively enhancing the original data by generating diverse yet semantically coherent samples. Such improvements enrich the feature representations of users and items, contributing to better recommendation performance.
Eliminating negative sampling leads to a substantial drop across all evaluation metrics. This highlights the module’s critical role in enhancing the model’s discriminative capability by providing high-quality hard negative samples, which complement the diffusion augmentation process.
Therefore, the experimental results demonstrate that both modules play pivotal roles in collaborative filtering and exhibit a synergistic superposition effect.
\begin{figure*}[htbp]
  \centering
  \begin{minipage}{0.325\textwidth} 
    \centering
    \includegraphics[width=\linewidth]{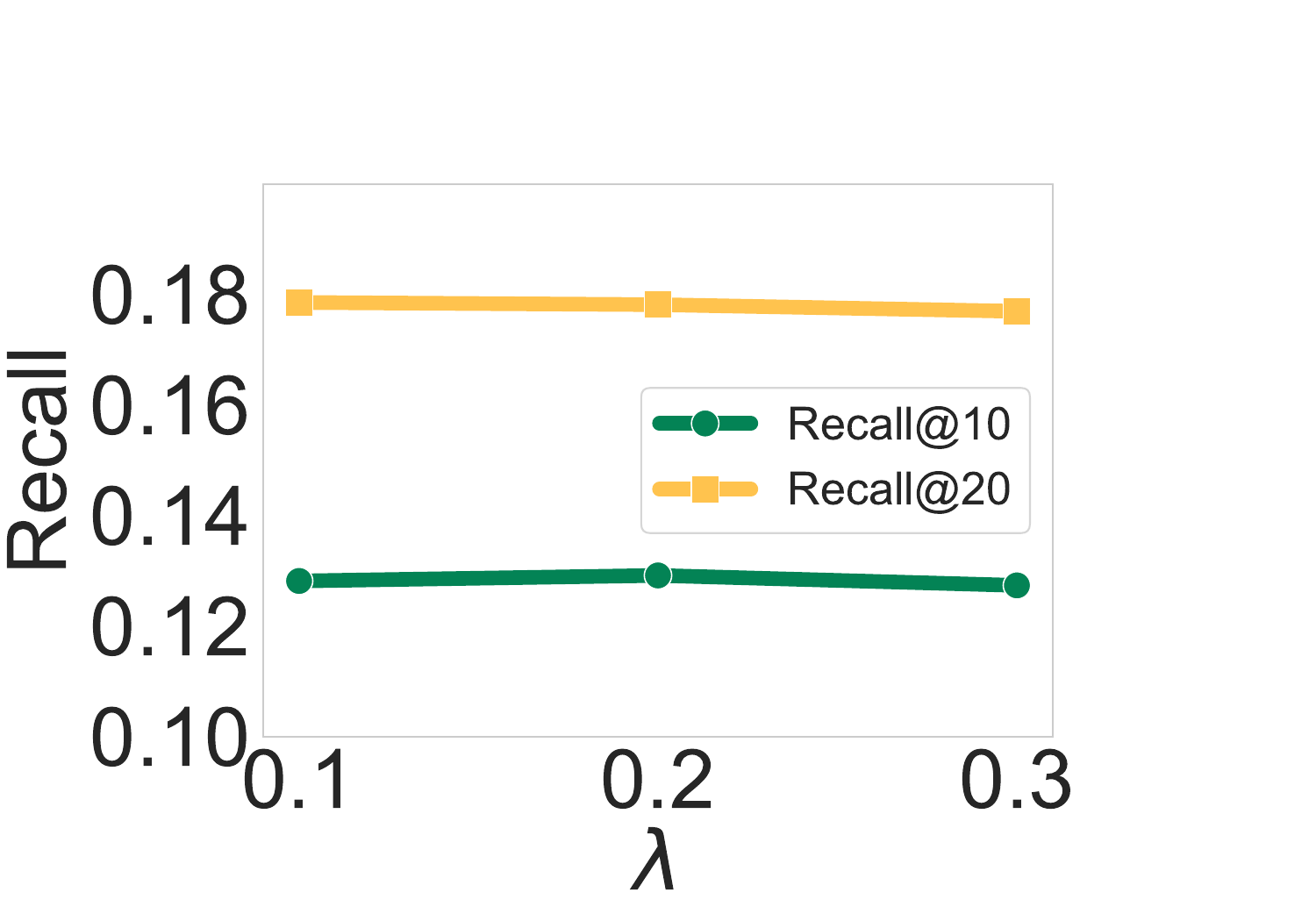}
    \caption*{(a)}
    \label{fig:a}
  \end{minipage}
  \hfill 
       \begin{minipage}{0.325\textwidth} 
    \centering
    \includegraphics[width=\linewidth]{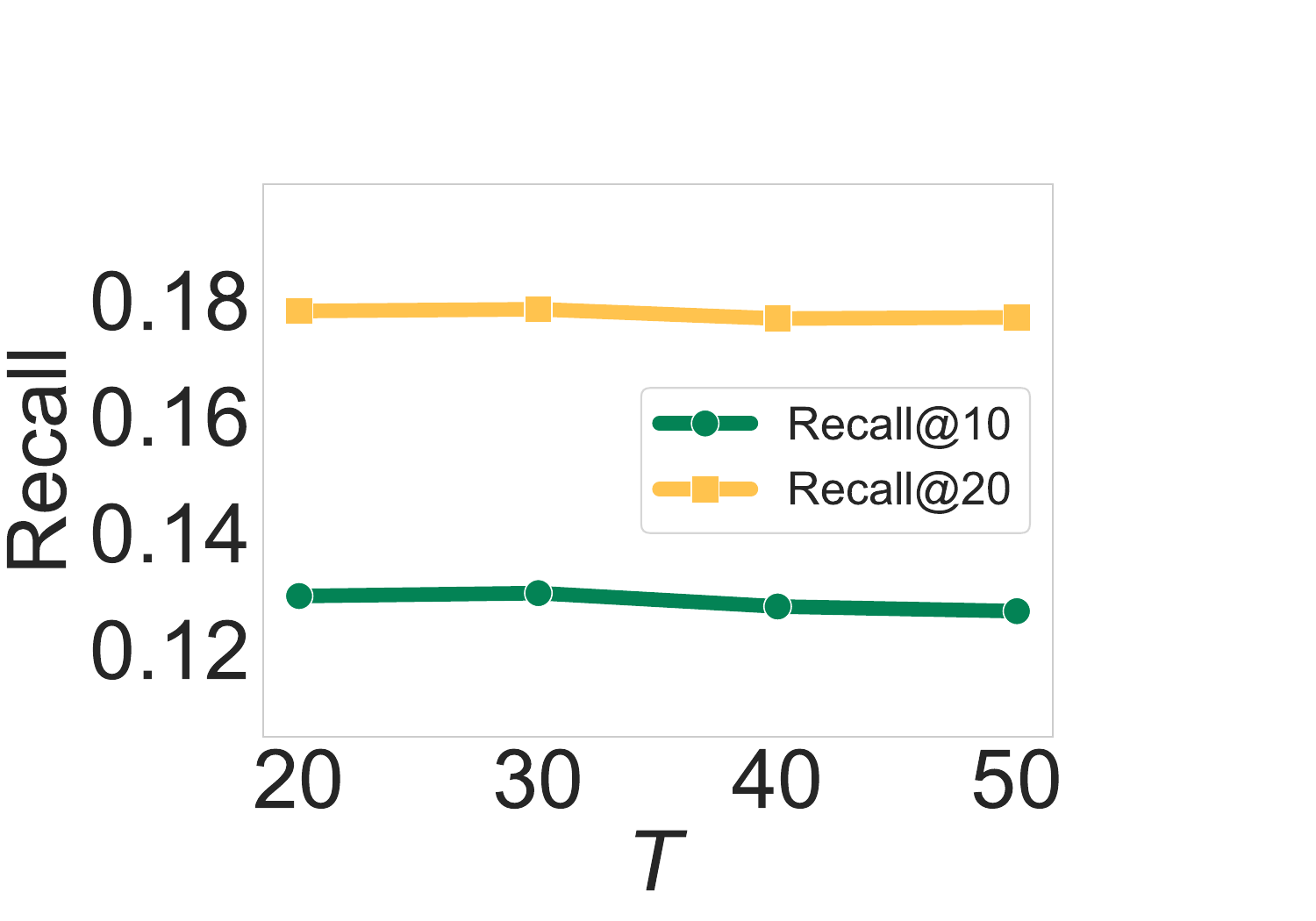}
    \caption*{(b)}
  \end{minipage}
  \hfill
  \begin{minipage}{0.325\textwidth} 
    \centering
    \includegraphics[width=\linewidth]{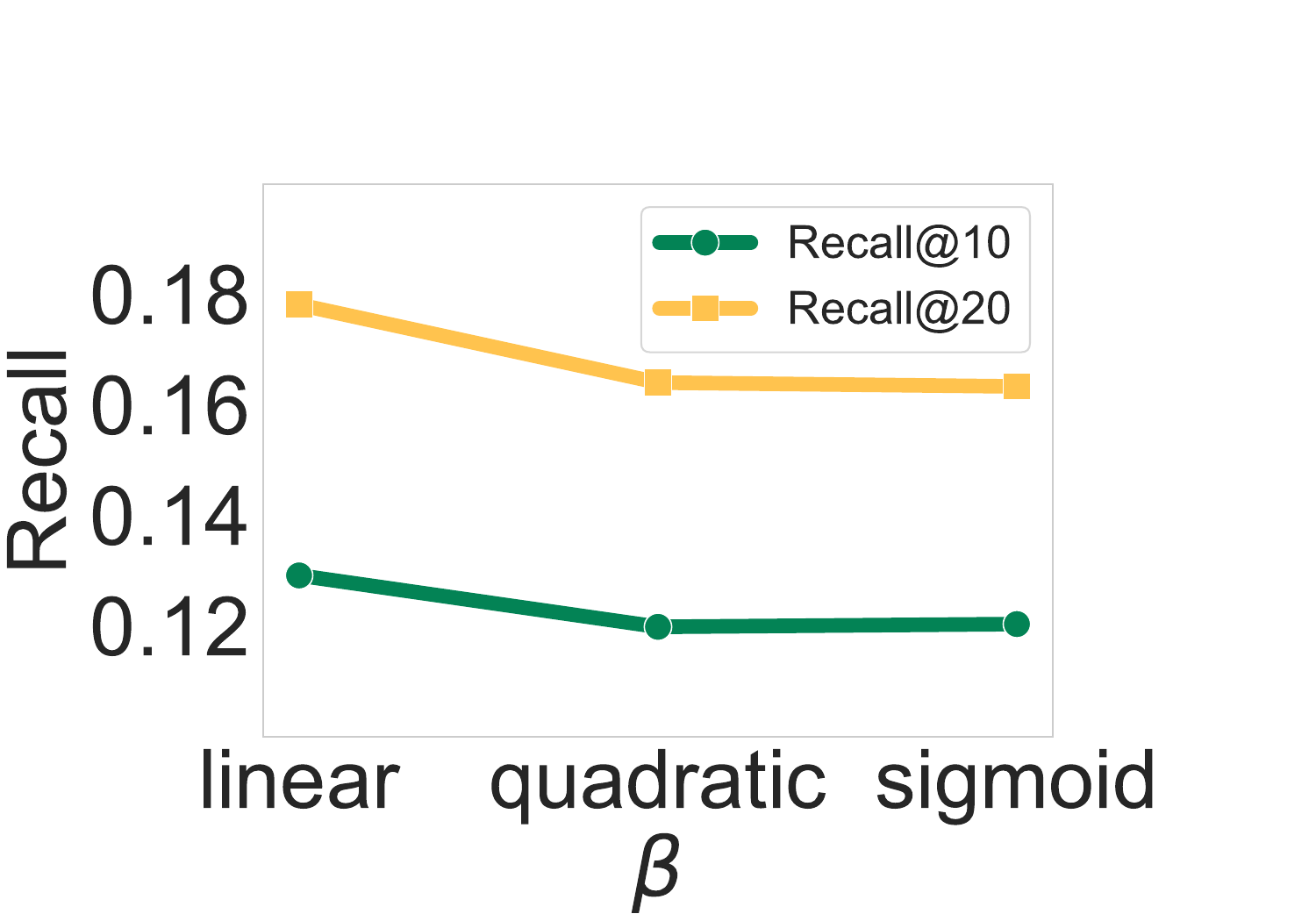}
    \caption*{(c)}
  \end{minipage}
  \hfill
\rule{\textwidth}{0pt}
  \begin{minipage}{0.325\textwidth}
    \centering
    \includegraphics[width=\linewidth]{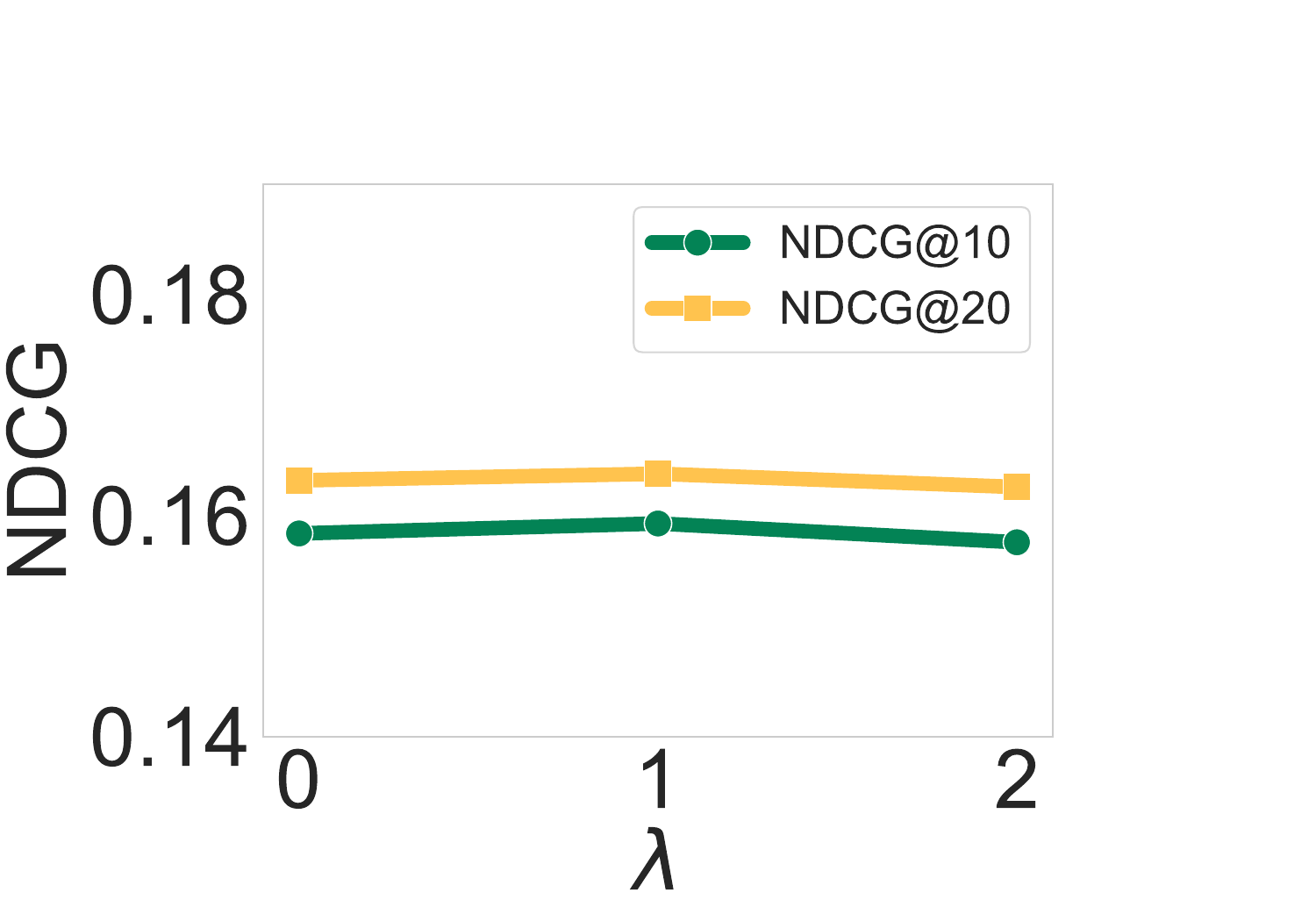}
    \caption*{(d)}
  \end{minipage}
  \hfill
  \begin{minipage}{0.325\textwidth}
    \centering
    \includegraphics[width=\linewidth]{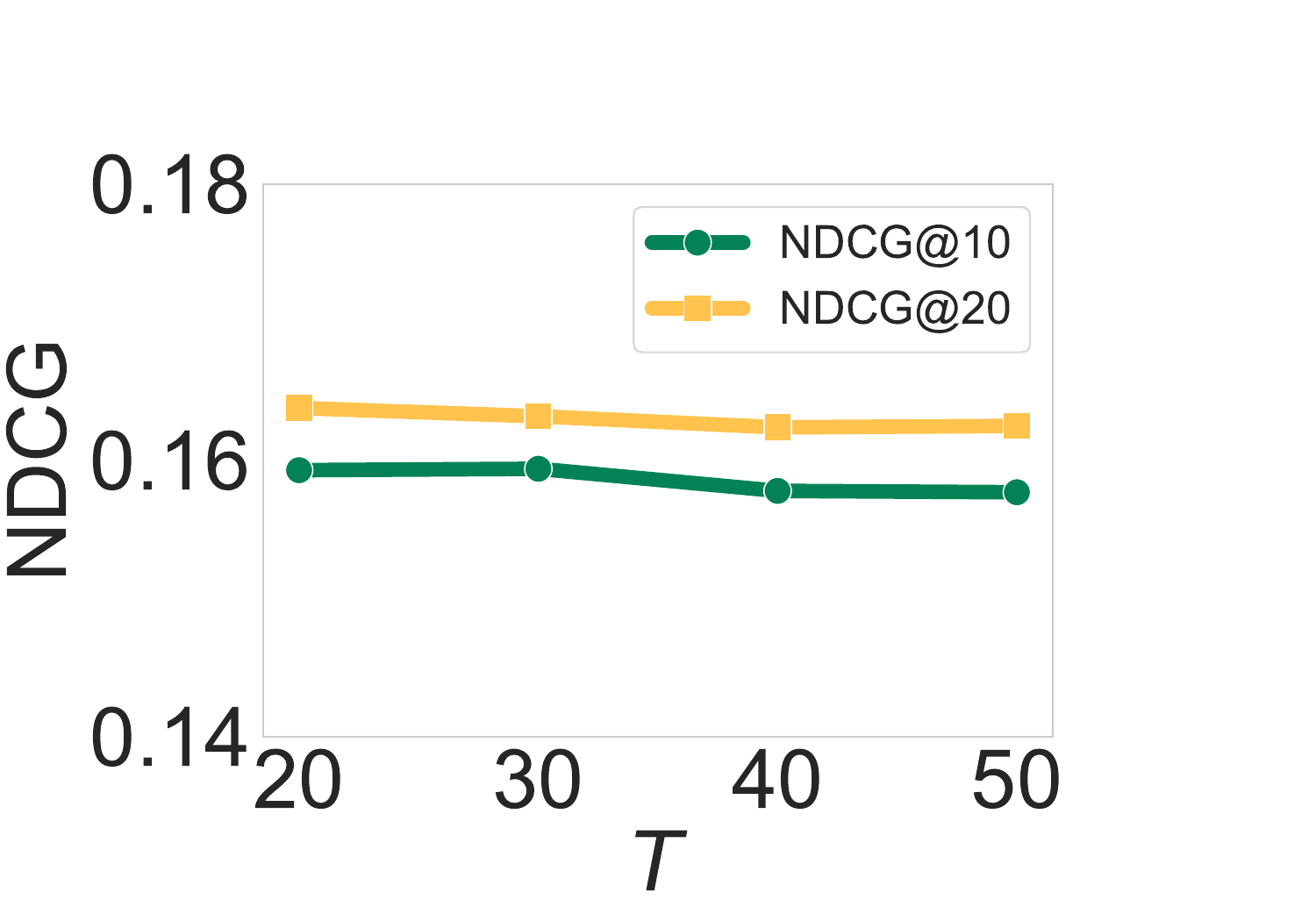}
    \caption*{(e)}
  \end{minipage}
  \hfill
  \begin{minipage}{0.325\textwidth}
    \centering
    \includegraphics[width=\linewidth]{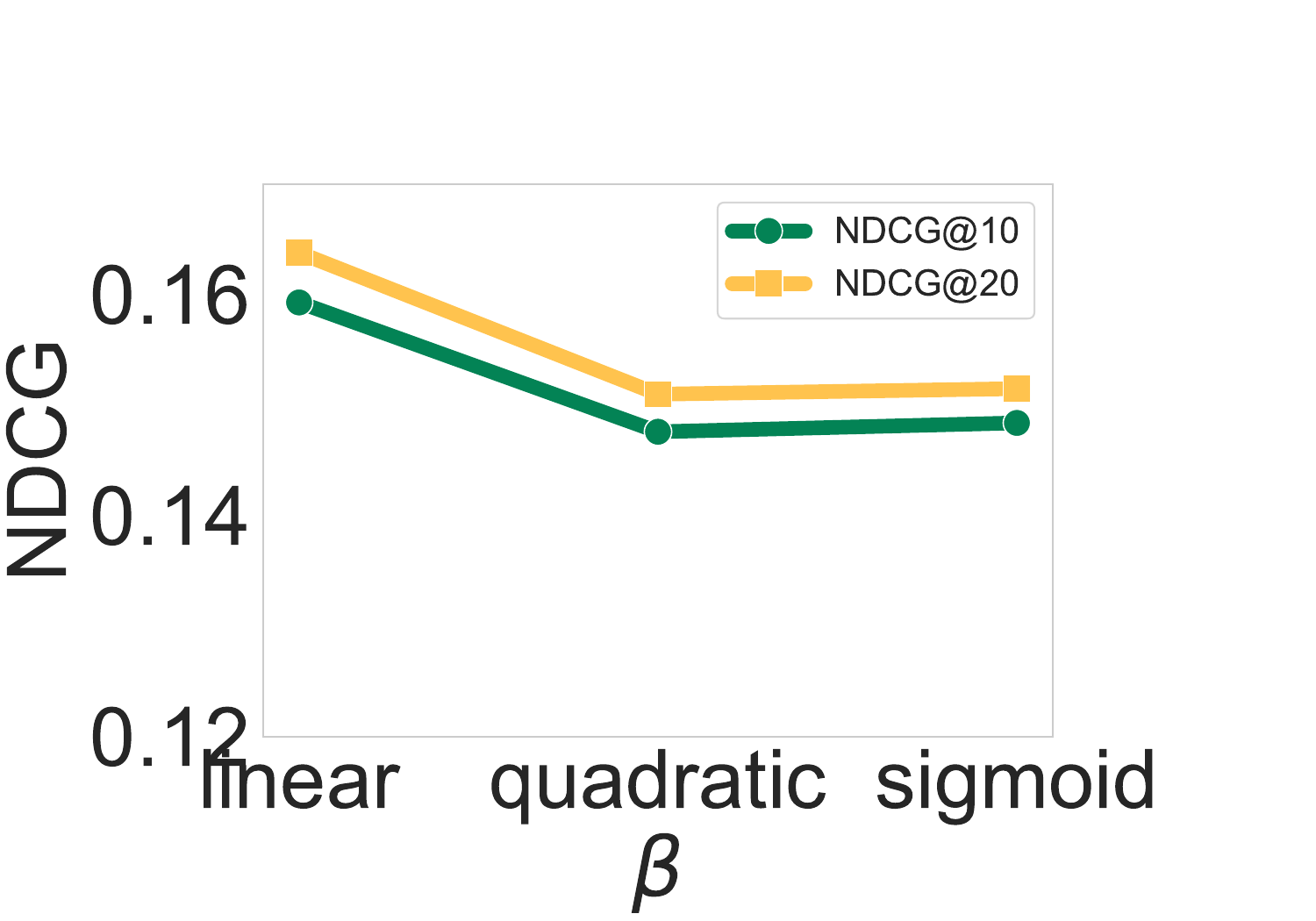}
    \caption*{(f)}
  \end{minipage}
\caption{Effect of the $\lambda$, diffusion step $T$ and noise $\beta$ on Douban-Book}
\label{parameter}
\end{figure*}
\begin{figure*}[h]
	\begin{minipage}{0.45\linewidth}
		\vspace{3pt}
		\centerline{\includegraphics[width=\textwidth]{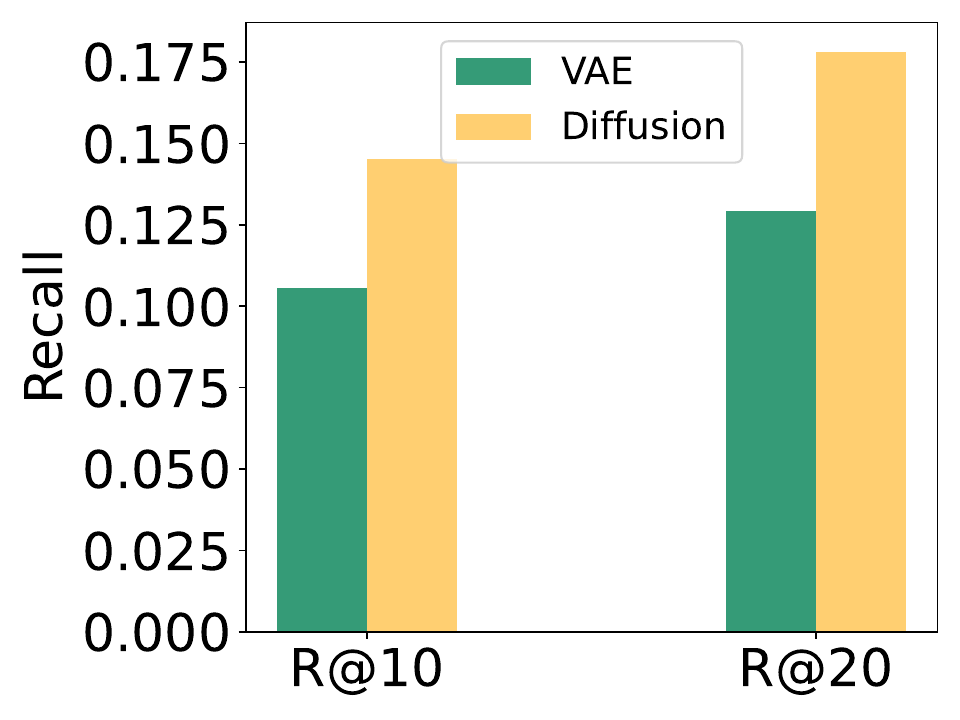}}
		\caption*{(a)}
	\end{minipage}
	\begin{minipage}{0.45\linewidth}
		\vspace{3pt}
		\centerline{\includegraphics[width=\textwidth]{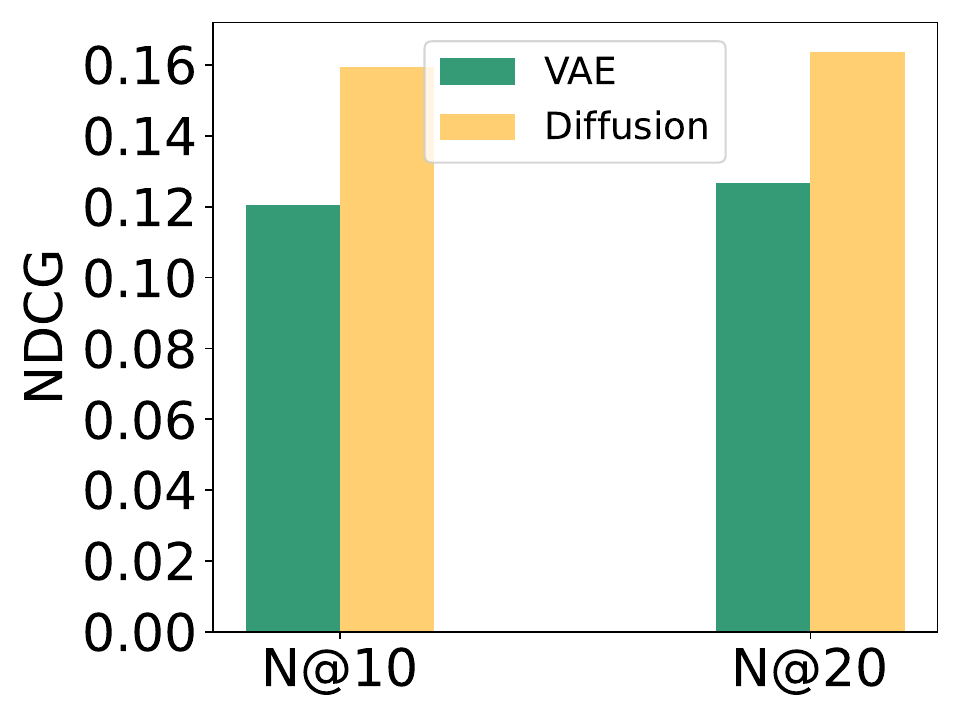}}
		\caption*{(b)}
	\end{minipage}
	\caption{Contrastive augmentation performance between Diffusion and VAE on Douban-Book.  }
	\label{case}
\end{figure*}
\subsection{Hyper-Parameter Sensitivity Analysis}
\noindent\textbf{Effect of Graph Layer Depth $L$.} To investigate the impact of graph layer depth, we vary $L$ within the range \{1,2,3\}. As shown in Table \ref{layer}, the model achieves optimal performance with $L=3$ across all three datasets. It is demonstrated the shallow graph layers can not capture the high-order neighbor interactions and semantic dependencies. 
We are not increasing the layers because too many layers risk over-smoothing in GNN learning, where node embeddings become indistinguishable due to excessive aggregation.

\noindent\textbf{Effect of Contrastive Loss $\lambda$.} As illustrated in Figure~\ref{parameter}(a),(d), We evaluate the influence of the contrastive loss weight $\lambda$ over\{0.1,0.2,0.3\} on the Douban-Book dataset. The results reveal that DGCL yields peak performance when $\lambda$ is 0.2. A balanced weighting $(\lambda=0.2) $ optimally integrates these objectives. High loss weight $(\lambda=0.3)$ overemphasizes contrastive regularization, diluting task-specific signals, while lower value $ (\lambda=0.1)$ underutilizes the benefits of contrastive learning. Therefore, the recommendation loss plays a dominant role and drives task-specific learning, and the contrastive enhancement loss function serves as an auxiliary component which enhances embedding robustness by promoting invariance to augmented contrastive views. The integration of these two loss functions and joint training can improve the performance of the recommendation task.

\noindent\textbf{Effect of Diffusion Step $T$.} The diffusion step $T$ critically governs the augmentation process by balancing noise injection and feature preservation. As illustrated in Figure~\ref{parameter}(b)(e), the results indicate that the metrics obtain superior results when $T$ is 30, with NDCG@20 slightly declining when $T$ is 30.
And as the diffusion step increases over time, the results tend to decrease gradually. This phenomenon may result from the multiple iterations of noise injection which may lead to excessive feature smoothing and the inability to capture the unique feature of each node. Moreover, more diffusion steps lead to more time cost and diversity loss. Therefore, sufficient steps are necessary to refine embeddings, but excess iterations harm discriminative power.

\noindent\textbf{Effect of the Noise Schedule $\beta$.} In this section, we evaluate three noise scheduling strategies for $\beta$: linear, quadratic, and sigmoid interpolation methods~\cite{a33}. As shown in Figure ~\ref{parameter}(c), (f), the linear schedule outperforms alternatives. This may attributed to that sable diffusion facilitates the balance of noise interference and semantic coherence. In contrast, non-linear schedules (quadratic, sigmoid) disrupt the balance between perturbation and stability, leading to suboptimal augmentation.
\subsection{Diffusion Augmentation Analysis }

To validate the superior augmentation capability of our diffusion-based approach, we conduct a comparative analysis using a Variational Autoencoder (VAE)-based generation. Specifically, we replace the diffusion module in DGCL with a VAE and evaluate both methods on the Douban-Book dataset. As illustrated in Figure \ref{case}, the diffusion-augmented method consistently outperforms its VAE-augmented counterpart.
This indicates that the multi-step denoising mechanism has advantages over VAE single-step reconstruction. The diffusion process employs iterative denoising steps to refine augmented samples progressively. Unlike VAE single-step augmentation, this gradual correction enables a deeper exploration of latent feature correlations, enhancing the model’s ability to capture complex user-item interactions.
Moreover, through implicit probabilistic modeling, the diffusion mechanism dynamically adjusts augmentation intensity based on local data density. DGCL can preserve semantic consistency with subtle perturbations in high-density regions and synthetic meaningful samples in low-density regions, thereby mitigating the data sparsity.

\section{Conclusion}
In this paper, we propose a novel Diffusion-augmented Graph Contrastive Learning model in collaborative filtering. It provides a new data augmentation that integrates the diffusion model with graph contrastive learning to enhance recommendation systems. Specifically, the diffusion process learns Gaussian distributions of node representations, generating augmented contrastive views by iteratively injecting and removing noise. 
The forward process injects the noise into the nodes' representations and the transformer-based encoder reverses the noise corruption, recovering discriminative embeddings while preserving topological relationships. 
The new contrastive views can be inferred from the estimated Gaussian distribution parameters. This augmented module also combines with a positive mixing of negative sampling and achieves the best performance on three public datasets. The results demonstrate that DGCL considers the semantic consistency in the feature space and node-specific features. In addition, it explores unrepresented regions of the feature space, thereby enriching the diversity of contrastive views. 

%
%
%
\bibliographystyle{splncs04}
\bibliography{reference.bib}
%




\end{document}